\documentclass[%
reprint,
superscriptaddress,
groupedaddress,
amsmath,amssymb,
aps,
]{revtex4-1}

\usepackage{graphicx}% Include figure files
\graphicspath{{figures/}} %Setting the graphicspath
\usepackage{dcolumn}% Align table columns on decimal point
\usepackage{bm}% bold math
\usepackage{times}% bold math
\newcommand{\bn}{{\bf n}}
\renewcommand{\bm}{{\bf m}}
\newcommand{\p}{{p}}
\newcommand{\brho}{\boldsymbol{\rho}}
\newcommand{\bsigma}{\boldsymbol{\sigma}}
\newcommand{\bpi}{\boldsymbol{\pi}}
\newcommand{\kl}{D_{\rm KL}}
\DeclareMathOperator*{\argmax}{argmax}

\begin{document}

%\preprint{APS/123-QED}

\title{Bayesian estimation of information-theoretic metrics for sparsely sampled distributions}

\author{Angelo Piga}
    \email{angelo.piga@gmail.com}
    \altaffiliation{Department of Chemical Engineering, Universitat Rovira i Virgili, Tarragona 43007, Catalonia.}
\author{Lluc Font-Pomarol}%
    \email{lluc.font@urv.cat}
    \altaffiliation{Department of Chemical Engineering, Universitat Rovira i Virgili, Tarragona 43007, Catalonia.}
\author{Marta Sales-Pardo}
    \email{marta.sales@urv.cat}
    \altaffiliation{Department of Chemical Engineering, Universitat Rovira i Virgili, Tarragona 43007, Catalonia.}
\author{Roger Guimerà}
    \email{roger.guimera@urv.cat}
    \altaffiliation{Department of Chemical Engineering, Universitat Rovira i Virgili, Tarragona 43007, Catalonia.}
    \altaffiliation{ICREA, Barcelona 08010, Catalonia.}

\date{\today}% It is always \today, today,


\begin{abstract}
Estimating the Shannon entropy of a discrete distribution from which we have only observed a small sample is challenging. Estimating other information-theoretic metrics, such as the Kullback-Leibler divergence between two sparsely sampled discrete distributions, is even harder. Existing approaches to address these problems have shortcomings: they are biased, heuristic, work only for some distributions, and/or cannot be applied to all information-theoretic metrics. Here, we propose a fast, semi-analytical estimator for sparsely sampled distributions that is efficient, precise, and general. Its derivation is grounded in probabilistic considerations and uses a hierarchical Bayesian approach to extract as much information as possible from the few observations available. Our approach provides estimates of the Shannon entropy with precision at least comparable to the state of the art, and most often better. It can also be used to obtain accurate estimates of any other information-theoretic metric, including the notoriously challenging Kullback-Leibler divergence. Here, again, our approach performs consistently better than existing estimators.

\end{abstract}

\maketitle

% \maketitle
% \doublespacing

\section{Introduction}

Information theory is gaining momentum as a methodological framework to study complex systems.
In network science, information theory provides rigorous tools to predict unobserved links~\cite{guimera09} and to infer community structure~\cite{peixoto-ent}. In neuroscience, Shannon entropy of spike train distributions characterizes brain activity from neural responses~\cite{bialek}, while mutual information identifies correlations between brain stimuli and responses~\cite{panzeri}. Recently, the Kullback-Leibler divergence~\cite{kullback1951} and its regularized version, the Jensen-Shannon distance, have also been successfully used in a wide variety of contexts: in cognitive science as a measure of ``surprise,'' to quantify and predict how human attention is oriented between changing screen images~\cite{itti}; in quantitative social science, in combination with topic models, to track the propagation of political and social discourses~\cite{dedeo, gerlach2016similarity} or to understand the emergence of social disruption from the analysis of judicial decisions \cite{font23}; and in machine learning, at the intersection between the statistical physics of diffusive processes, probabilistic models and deep neural networks~\cite{bahri2020kl}.

Information theoretical metrics are measured on distributions. In practice, a distribution $\brho$ over the possible states of a system, as well as functions $\mathcal{F}(\brho)$ of this distribution (such as Shannon entropy or other metrics), have to be inferred from experimental observations. However, this inference process is difficult for many real complex systems since, due to experimental limitations, the observations are often sparse, and statistical estimates of the distribution $\brho$ and its functions can be severely biased~\cite{subsampling-rev}. Here, we focus on the particular yet important case of discrete (or categorical) distributions $\rho_i$, $i=1,\dots, K$, where $K$ is the number of possible states (or categories), which is known and fixed. Inferences about $\brho$ and any function must be based on $n_i$, the number of observations in the $i$-th state (with $N = \sum_i n_i$ the sample size) and, in the undersampled regime we are interested in, $N \lesssim K$. The challenge is thus, from the sparse observations $\{n_i\}$, to infer the probability $\rho_i$ of each category $i$ and estimate metrics $\mathcal{F}(\brho)$.

A theoretically well-founded approach to tackle this problem is provided by the principles of conditional probability, encapsulated in Bayes' theorem~\cite{jaynesbook}. This framework is in general preferable because of its transparency---it requires that all assumptions of the underlying generative model for the data are made explicit, expressed via the choice of a likelihood function and a prior distribution that reflects the knowledge about the system before observing any data. In probabilistic reasoning, the combination of observations and prior distribution provides an updated (posterior) probability distribution of the quantity under study. 
%, that is an estimation of its mean value together higher order moments. 
%In particular it can be proven that the Bayes estimator has the minimum variance~\cite{wolpert1}. 
Other estimation strategies make implicit assumptions and often provide only point estimates, as opposed to full distributions. 

A class of expressive generative models for categorical distributions amenable to a Bayesian framework is the well-studied family of Dirichlet distributions. However, as Nemenmann, Shafee, and Bialek (henceforth NSB) pointed out in ~\cite{nemenman2001}, when sample sizes are small ($N \lesssim K$), the Shannon entropy inferred from assuming that the observed distribution is sampled from the Dirichlet family is tightly determined by the specific parameters one chooses for the Dirichlet model; therefore, inaccurate choices result in severe biases of the Shannon entropy estimates. To overcome this problem, they introduced a mixture of Dirichlet models, which results in a very precise estimator of the Shannon entropy that works for a wide variety of distributions, even in the sparse sampling regime $N \lesssim K$ \cite{nemenman2001,nemenman2004}.

Although, in terms of precision, NSB can be considered the state of the art for estimating the Shannon entropy, it does not provide estimates for the distribution $\brho$. For this reason, its applicability is limited to estimating the Shannon entropy (and related information theoretic quantities like mutual information and Jensen-Shannon distance, which can be expressed in terms of entropies). By contrast,  it cannot be used to estimate the Kullback-Leibler divergence.
To cover this gap,  Hausser and Strimmer (henceforth HS) derived a James-Stein-type shrinkage estimator for $\brho$~\cite{hausser}, which has the advantage of being analytical and applicable to any information-theoretic metric, but at the price of making implicit {\em ad hoc} assumptions, of being less precise than NSB for the Shannon entropy, and of lacking error estimation. 

Here, we propose an alternative semi-analytical estimator for distributions that is efficient, precise, and general. Its derivation is grounded in probabilistic considerations, without any {\em ad hoc} assumptions. We consider Dirichlet generative models and use a hierarchical Bayesian approach to extract as much information as possible from the few observations at hand. In the case of Shannon entropy, we can estimate the expected value and higher order moments with precision at least comparable to the NSB estimator, and most often better. Additionally, because our method provides estimates of the probability distribution, it can be used to obtain accurate estimations of the Kullback-Leibler divergence. In this case our approach also performs equally or better than existing estimators.

\section{Background}\label{sec.background}

Let us consider a system with $K$ possible output states whose observations follow an unknown discrete distribution ${\brho=\{\rho_i;\, i=1,\dots,K\}}$ with ${\sum_i \rho_i=1}$. The vector ${\bn=\{n_i;\, i=1,\dots,K\}}$ represents the number of times each state was observed in a set of ${\sum_i n_i=N}$ independent observations of the system. We also consider a function $\mathcal{F}(\brho)$ of $\brho$, such as, for example, the Shannon entropy 
\begin{equation}\label{eq.shannon}
S(\brho) = - \sum_{i=1}^K \rho_i \log \rho_i\;,
\end{equation}
which we want to estimate from the set of observations.

\begin{figure*}[t]
  \centering
  \includegraphics[width=2\columnwidth]{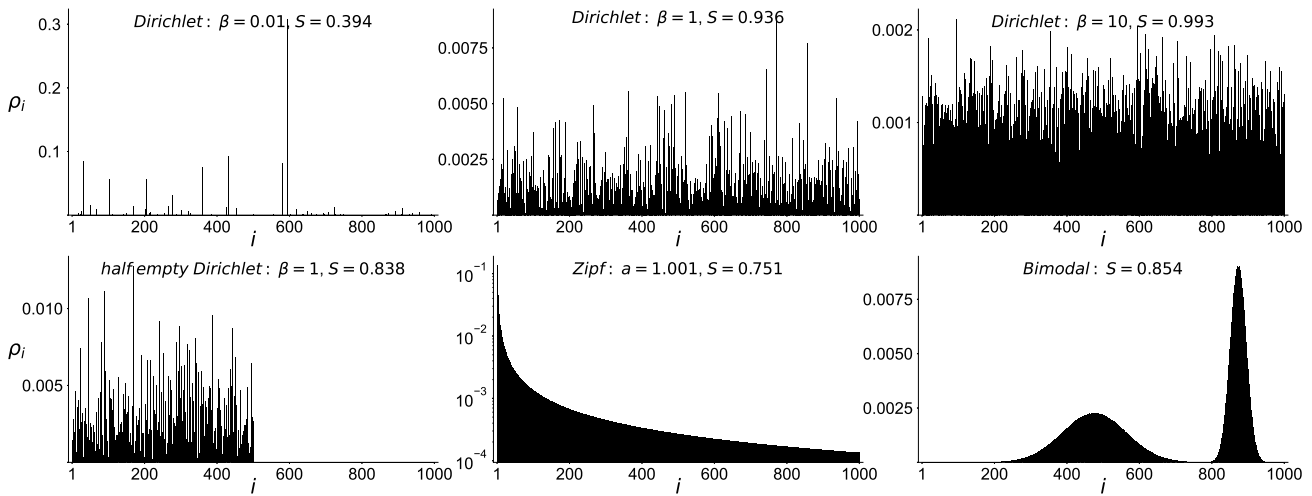}
  \caption{Examples of target distributions. First row: three categorical distributions sampled from uniform Dirichlet with $\beta = 0.01, 1, 10$, respectively. Second row: a categorical distribution sampled from a uniform Dirichlet, $\beta=1$, but where half bins are set to zero; Zipf's distribution with exponent $a=1.001$; bimodal distribution: two gaussians with $\{ \text{mean},\text{standard deviation}\}$ respectively $\{10,20\}$ and $\{100,5 \}$, are concatenated and then discretized over a histogram of $1000$ categories.}
  \label{fig.dist}
\end{figure*}

The posterior distribution over the values of the function $\mathcal{F}$ given the observed counts $\bn$ is
\begin{equation}\label{eq.function}
   p\left(\mathcal{F}|\bn \right) = \int d\brho \; \delta \left( \mathcal{F} - \mathcal{F}(\brho)\right) p(\brho |\bn) \;, 
\end{equation}
where $p(\brho |\bn)$ is the posterior of the distribution $\brho$ given the counts $\bn$. We further assume that the prior over distributions depends on a parameter $\beta$, which becomes a hyperparameter of our generative model. 
Then, using the laws of conditional probability, we can write the posterior $p(\brho |\bn,\beta)$  as
\begin{equation}\label{eq.bayesrules}
    p(\brho|\bn, \beta) = \frac{p(\bn |\brho, \beta) \, \p(\brho|\beta)}{p(\bn|\beta)} \;,
\end{equation}
where $p(\bn|\brho, \beta)$ is the likelihood, $\p(\brho|\beta)$ is the prior over distributions, 
and $p(\bn|\beta) = \int d\brho\, p(\bn|\brho)\, \p(\brho|\beta)$ is the evidence and acts as normalization factor.
%Everywhere, $\beta$ parametrizes the prior over distributions and is thus a hyperparameter of the generative model.
The likelihood is the probability of the empirical observations $\bn$ given $\brho$; for independent multinomial samples, the probability of observing an event of type $i$ is $\rho_i$, and the full likelihood is the product $p(\bn|\brho, \beta) = p(\bn|\brho) = N! \prod_i^K\rho_i^{n_i}/n_i!$ and, given $\brho$, it is independent of the hyperparameter $\beta$. The prior $\p(\brho|\beta)$  expresses the probability of each distribution $\brho$ prior to observing any data, and plays a crucial role in the discussion below.
Symmetric Dirichlet distributions are convenient priors because they are a generative model for a broad class of discrete distributions. Additionally, they have been widely used in this setting \cite{gelmanbook}; they are parametrized as follows
\begin{equation}\label{eq.prior}
    \p(\brho|\beta) = \frac{1}{{\rm B}_K(\beta)} \prod_{i=1}^K \rho_i^{\beta-1}\;, \quad \mathrm {B}_K(\beta)=\frac{\Gamma(\beta)^{K}}{\Gamma(\beta K)}\;,
\end{equation}
where $\Gamma$ is the gamma function, while the hyperparameter $\beta$ is a real, positive number known as the concentration parameter. In the first row of Fig.~\ref{fig.dist}, examples of categorical distributions sampled from symmetric Dirichlet priors are shown. 

Besides being very expressive, Dirichlet priors are conjugate distributions of categorical likelihoods, meaning that the posterior is still a Dirichlet distribution, a property that often makes the inference via Eqs.~\eqref{eq.bayesrules} and~\eqref{eq.function} analytically tractable. For example, when $\mathcal{F}(\brho) = \brho$, Dirichlet priors lead to expected posterior probabilities $\langle \rho_i \rangle$ given by the widely-used generalized Laplace's formula
\begin{equation}\label{eq.laplace}
    \langle \rho_i\rangle = \frac{n_i+\beta}{N+K\beta}\;.
\end{equation}
It is worth noting the improvement of Eq.~\eqref{eq.laplace} with respect to the maximum likelihood (or frequency) estimator $\rho_i = n_i/N$, which is recovered by the former in the limit $\beta \rightarrow 0$. In particular, Laplace's formula assigns non zero probability to non observed states, a desirable property whose advantage will become evident later, when estimating Kullback-Leibler divergences. This example also illustrates how non-Bayesian approaches to inference make implicit and non-trivial assumptions, in this case  assuming $\beta\rightarrow 0$ amounts to assuming that infinitely concentrated distributions $\brho$ are a priori much more plausible than more homogeneous ones.

Going back to the estimation of $\mathcal{F}$ from the observations $\bn$, and given Eq.~\eqref{eq.laplace}, one may be tempted to directly plug the value of $\langle \rho_i \rangle$ in the explicit expression of $\mathcal{F}(\brho)$ to get a point estimate. However, this is just an approximation;
the exact procedure consists in finding and using the whole posterior $p(\mathcal{F}|\bn)$. Specifically, the expected value of this posterior $\langle \mathcal{F} \rangle = \int d\mathcal{F} \, \mathcal{F} \, p(\mathcal{F}|\bn)$ minimizes the mean-squared error~\cite{wolpert1}, and its mode is a consistent estimator, meaning that it converges to the true value of $\mathcal{F}(\brho)$ when the number of observations increases, regardless of the prior and, in particular, regardless of the hyperparameter $\beta$. Wolpert and Wolf in Refs.~\cite{wolpert1,wolpert2} provided analytical formulas for all the moments of $p(\mathcal{F}|\bn)$ when $\mathcal{F}$ is the Shannon entropy and for Dirichlet priors (we report the formula for the mean in Eq.~\eqref{eq.wwmean} and for the second moment in Appendix~\ref{ap:moments}).

However, an unbiased estimation of $\mathcal{F}$ is not guaranteed for small samples. This is often the case for Dirichlet priors, especially when the parameter $\beta$ is unknown.
Several options for the value of $\beta$ have been proposed in literature, each one suitable to some specific case but deficient in others (for a discussion, refer to Refs.~\cite{nemenman2001, hausser}). 
In \cite{nemenman2001}, NSB suggested that, when samples are scarce, any attempt to find a single universal $\beta$ is hopeless; the fundamental reason being that categorical distributions generated by a Dirichlet have a Shannon entropy that is narrowly determined by, and monotonically dependent on, $\beta$. In other words, for small samples, the posterior distribution~\eqref{eq.function} is dominated by the prior. To overcome this problem, Refs.~\cite{nemenman2001, nemenman2004} proposed, as the prior $\p_{\rm NSB}(\brho)$, an infinite mixture of Dirichlet priors
\begin{equation}\label{eq.nsbprior}
    %\mathcal{P}_{\rm NSB}(\brho) \propto \int d\beta \frac{d\overline{S}(\beta ; K)}{d\beta} \mathcal{P}(\brho|\beta)\;, 
    \p_{\rm NSB}(\brho) \propto \int d\beta\, \p_{\rm NSB}(\beta)\,\p(\brho|\beta)\;,
\end{equation}
where the weights $\p_{\rm NSB}(\beta)$ %(that is, the hyperpriors over $\beta$)
were set so as to obtain a flat prior over entropies $S$, and have the functional form
\begin{equation}\label{eq.nsb-hyp}
    \p_{\rm NSB}(\beta) \propto \frac{d\,\mathbb{E}[S|n_i=0,\beta]}{d\beta} = K\psi_1(K\beta + 1) - \psi_1(\beta + 1)\;,
\end{equation}
where $\mathbb{E}[S|\bn,\beta]$ is the expected entropy given the observations $\bn$, and then $\mathbb{E}[S|n_i=0,\beta]$ is the expected entropy of the distributions $\brho$ generated from a symmetric Dirichlet priors (that is if there are no observations), with fixed $\beta$ and $K$, and $\psi_m(x)=\left(\frac{d}{dx}\right)^{m+1} \log \Gamma(x)$ are the polygamma functions. 
The NSB prior leads to very accurate estimates of the Shannon entropy, and can be considered the state of the art. 
Even if best suited for situations in which the number of states $K$ is known and fixed, it is quite versatile and has been later extended for countable infinite number of states~\cite{archercountable} and further optimized for binary states~\cite{archerbinary} and long tail distributions~\cite{archercountable}. Other estimators, for example, the Chao-Shen estimator~\cite{chaoshen}, perform at most as well as the NSB (or its derivatives), but never better (see~\cite{hausser} for a comprehensive review).
Additionally, given an estimator of $S$, a number of other quantities can be indirectly estimated. For example, the mutual information $M$ between two distributions $\brho$ and $\bsigma$ is $M(\boldsymbol{\rho} \;;\boldsymbol{\sigma}) = S(\boldsymbol{\rho}) + S(\boldsymbol{\sigma}) - S(\bpi)$, where $\boldsymbol{\pi}$ is the joint distribution of $\boldsymbol{\rho}$ and $\boldsymbol{\sigma}$ ~\cite{archermutual}. Similar relations can be derived for Jensen-Shannon distance and other information-theoretic quantities~\cite{dedeo2013bootstrap}~\footnote{As observed in ~\cite{archermutual} and~\cite{wolpertdedeo}, mutual information can be expressed in terms of different combinations of the Shannon entropy of the two distributions. But its estimations in general differ. The expression $M(\rho \;;\sigma) = S(\rho) + S(\sigma) - S(\pi)$ seems to be the less biased, however, in the absence of a unique consistent prior over the joint distribution, it is not guaranteed it minimizes the mean-squared error.}.

However, consider the estimation of the Kullback-Leibler divergence ($\kl $) between two distributions $\brho$ and $\bsigma$ with the same dimension $K$
\begin{equation}\label{eq.kl}
	\kl (\brho\|\bsigma ) = \sum_{i=1}^K \rho_i \log_2 \frac{\rho_i}{\sigma_i}\;.
\end{equation}
To estimate $\kl $ from samples ${\bn=\{n_i;\, i=1,\dots,K\}}$ from $\brho$, and ${\bm=\{m_i;\, i=1,\dots,K\}}$ from $\bsigma$, one cannot use the NSB approach. First, $\kl $ is not a combination of the Shannon entropies of the two underlying distributions $\brho$ and $\bsigma$. Second, $\kl $ is unbounded, and any attempt to find a hyperprior in the spirit of Eq.~\eqref{eq.nsb-hyp} results in improper hyperpriors. Finally, with the NSB prior one renounces to any estimation of $\beta$ and, therefore, to a good a point estimation of $\kl $ by means of Laplace's formula. 

\begin{figure}[t]
     %%\begin{subfigure}[b]{1\columnwidth}
         \includegraphics[width=\columnwidth]{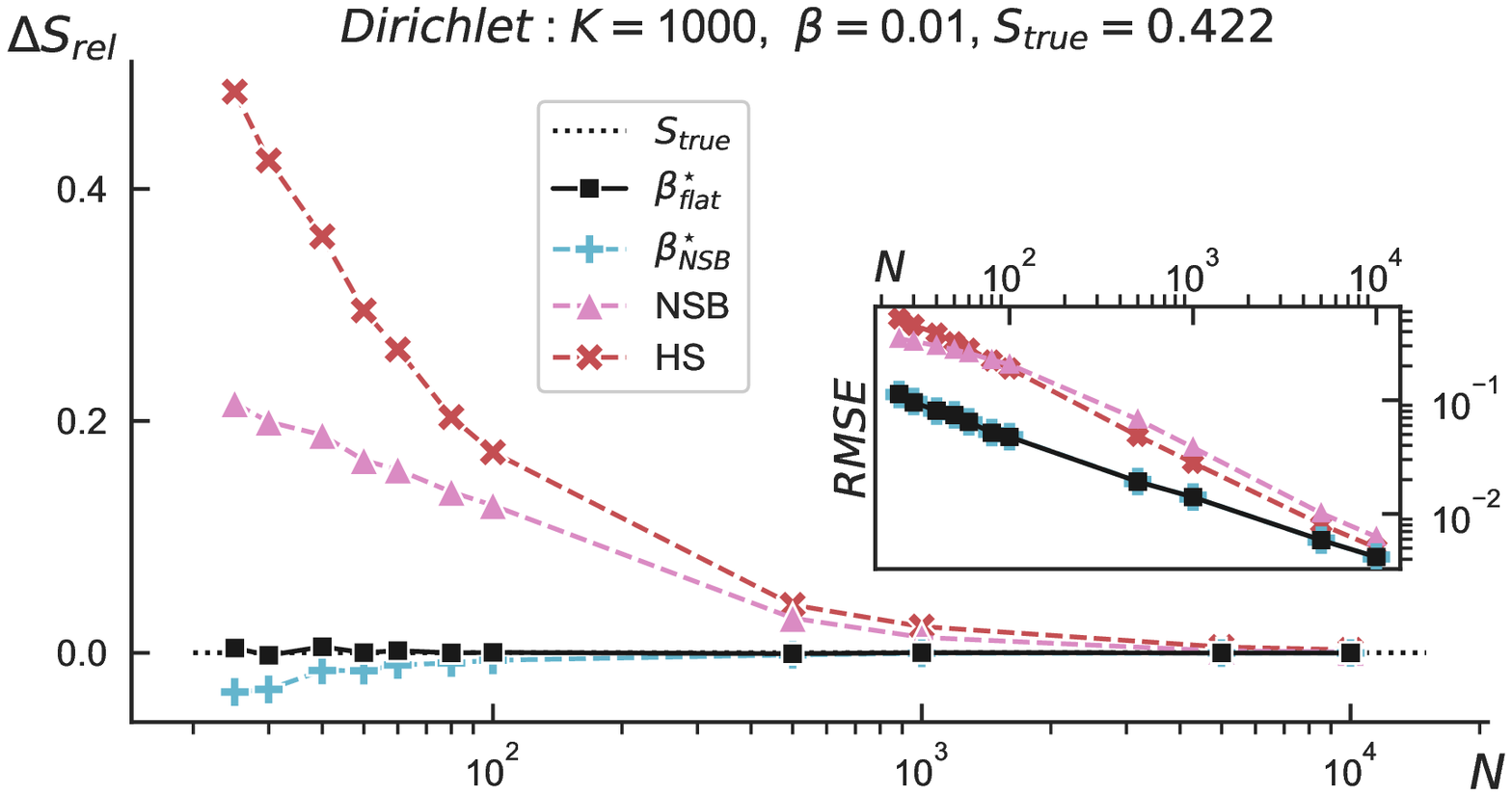}
        %  \caption{$y=x$}
     %%\end{subfigure}
     \vfill
     %%\begin{subfigure}[b]{1\columnwidth}
         \includegraphics[width=\columnwidth]{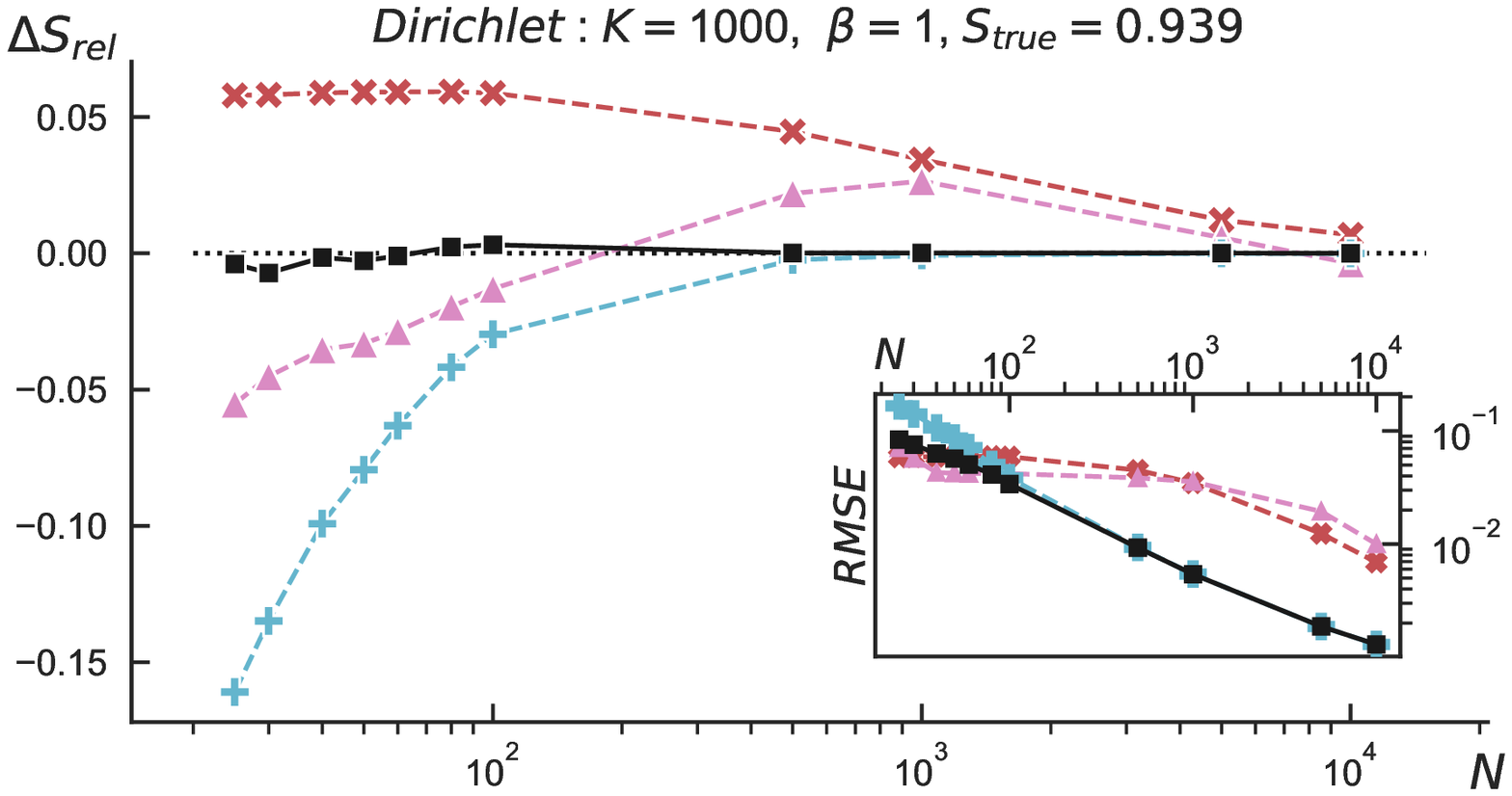}
     %%\end{subfigure}
     \vfill
     %%\begin{subfigure}[b]{1\columnwidth}
         \includegraphics[width=\columnwidth]{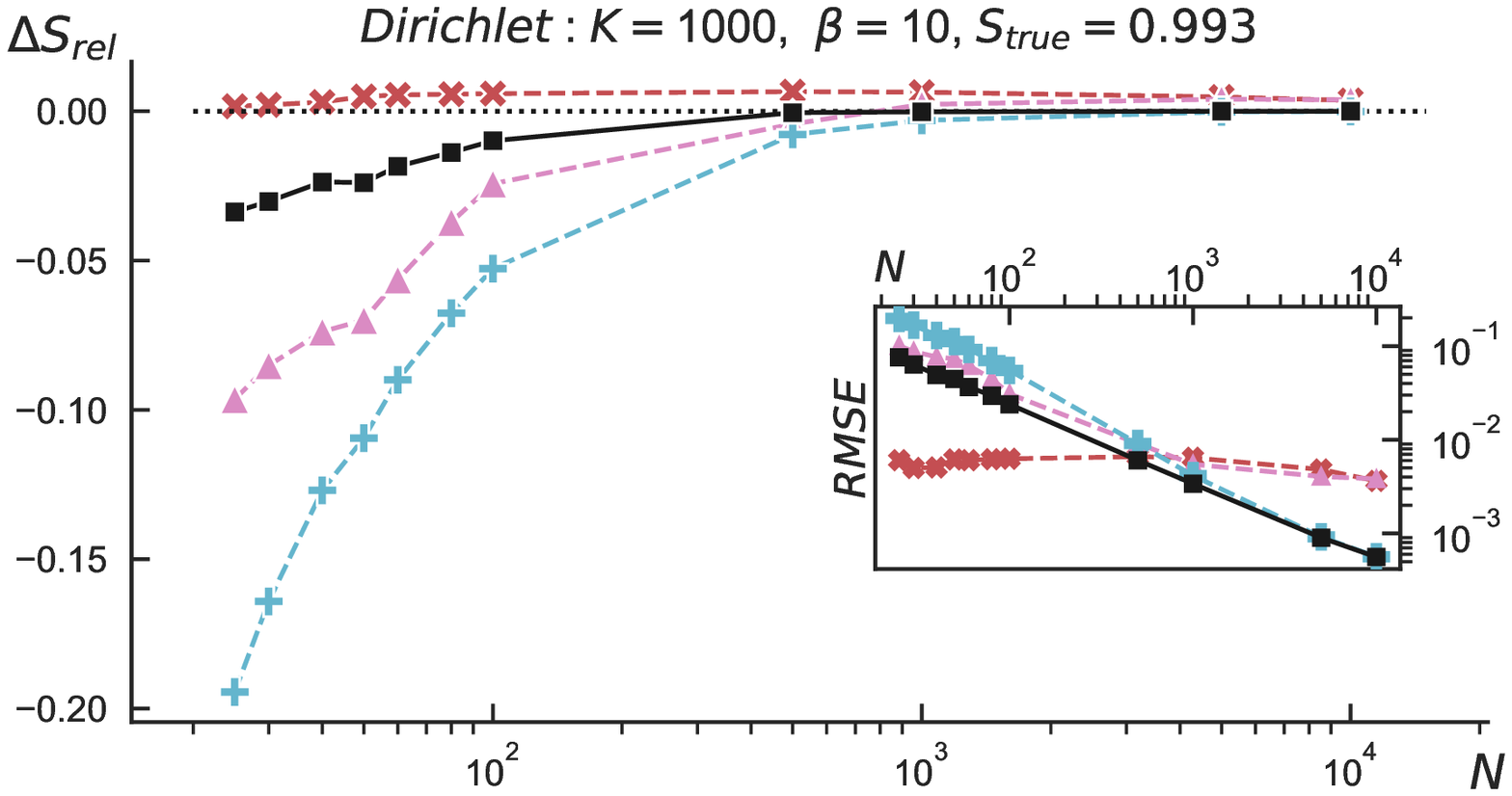}
        %  \caption{$y=5/x$}
     %%\end{subfigure} 
    \caption{Shannon entropy estimation for distributions typical in a Dirichlet prior, for $\beta=0.01, 1, 10$ and sample size $N = 25,\ldots , 10000$. Each point corresponds to an average over $1000$ samples. Main plots: relative errors of entropies $\Delta S_{rel}=(S_{est}-S_{true})/S_{true}$, where $S_{est}$ is the estimated entropy with different methods. Black squares: our estimator with $\beta^{\star}$ from a flat hyperprior. Cyan pluses: our estimator but with $\beta^{\star}$ from NSB hyperprior. Pink upper triangle: NSB estimator. Red crosses: Hausser-Strimmer estimator. Insets: roots mean-squared errors (note the logarithmic scale in both axes). The value of $S_{true}$ in the titles serves as a reference and indicates the average over the entropies of the runs. Standard-errors bars of the main plots are smaller then symbols and are not shown.}
        \label{fig:S_typ}
\end{figure}

\section{Hierarchical Bayes point estimate for $\beta$}

Here, we address these limitations of the NSB estimator while maintaining and even improving its performance. We posit that the success of the NSB approach stems, not from mixing infinitely many values of the concentration parameter $\beta$, but rather from the flexibility to accommodate for {\em any particular value} of $\beta$. Indeed, we surmise that, in general, only a narrow interval of $\beta$ values are compatible with a given observation $\bn$ and therefore contribute to the mixture, whereas most others do not contribute. Motivated by this, we propose an approach that aims to directly estimate the value of $\beta$ that most contributes to the posterior given the data $\bn$.

First, we observe that the posterior $p(\brho|\bn)$ can be written as
\begin{equation}
\begin{split}
    p(\brho|\bn) & = \int d\beta \, p(\brho|\bn, \beta) \, p(\beta|\bn) \\
    %& = \int d\beta \frac{p(\bn |\brho, \beta) \, \mathcal{P}(\brho|\beta)}{p(\bn|\beta)} \, p(\beta|\bn)\\
    & = \int d\beta \frac{p(\bn |\brho) \, \p(\brho|\beta)}{p(\bn|\beta)} \, p(\beta|\bn) \;,
\end{split}
\end{equation}
%
% The first equality is a direct application of the rules of the conditional probability. 
where we have applied Bayes' rule, and the fact that $\bn$ conditioned on $\brho$ is independent of $\beta$,  so that $p(\bn |\brho, \beta) = p(\bn |\rho)$.

Then, we assume that the conditional distribution $p(\beta|\bn)$ is very peaked around a given value $\beta^{\star}$, so that the posterior $p(\brho|\bn)$ can be approximated as
\begin{equation}\label{eq.post-approx}
    p(\brho|\bn) \approx \frac{p(\bn |\brho) \, \p(\brho|\beta^{\star})}{p(\bn|\beta^{\star})} \;.
\end{equation}
This approximation, sometimes referred to as {\em empirical Bayes}, is a point estimate for the fully hierarchical probabilistic model given by $p(\bn|\brho)$ and $\p(\brho|\beta)$. Eq.~\eqref{eq.post-approx} is identical to Eq.~\eqref{eq.bayesrules}, with the difference that the concentration parameter is now the most likely value of $\beta$ given the observed counts $\bn$, that is,
\begin{equation}
    \beta^{\star} = \argmax_{\beta}\, p(\beta|\bn) = \argmax_{\beta}\,\frac{p(\bn|\beta)\,\p(\beta)}{p(\bn)}~,
    \label{eq.marginal}
\end{equation}
where $p(\bn|\beta)=\int d\brho\, p(\bn|\beta, \brho) p(\brho|\beta)$.
For Dirichlet priors as in Eq.~\eqref{eq.prior}, $\beta^*$ satisfies (see Appendix~\ref{sec.supp})
\begin{equation}\label{eq.result}
    \sum_{i=1}^{K}\sum_{m=0}^{n_i-1} \frac{1}{m+\beta^{\star}}-\sum_{m=0}^{N-1}\frac{K}{m+K\beta^{\star}} + \frac{1}{\p(\beta^{\star})}\frac{d\, \p(\beta)}{d\,\beta}\Big|_{\beta^{\star}} = 0 ~,
\end{equation}
which is the key analytical result of this paper.

The hyperprior $\p(\beta)$ reflects our prior knowledge about the shape of the distribution of the hyperparameter. To be completely agnostic in this regard, we can use a uniform hyperprior
\begin{equation}\label{eq.flat-hyp}
    \p_U(\beta) = \frac{1}{\Delta\beta} = \text{const.}\;, \quad \Delta\beta = \beta_{\rm max} - \beta_{\rm min}\;,
\end{equation}
with cut-offs $0<\beta_{\rm min}<\beta_{\rm max} <\infty$. In this case, the derivative term in Eq.~\eqref{eq.result} disappears. The NSB hyperprior~\eqref{eq.nsb-hyp} is a valid alternative; in this case, the last term in Eq.~\eqref{eq.result} is (see appendix A for details)
\begin{equation}\label{eq.nsb-factor}
\frac{1}{\p_{NSB}(\beta^*)}\left. \frac{d\ \p_{\rm NSB}(\beta)}{d\,\beta}\right|_{\beta^*} = \frac{K^2\psi_2(k\beta^{\star}+1)-\psi_2(\beta^{\star}+1)}{K\psi_1(k\beta^{\star}+1)-\psi_1(\beta^{\star}+1)}\;.
\end{equation}

Despite the complex appearance of Eq.~\eqref{eq.result}, $\beta^*$ is not hard to obtain numerically, giving a computational improvement with respect the NSB estimator, whose algorithm is involved and has higher computational costs \footnote{Although we have not proved that the solution $\beta^{\star}$ is unique, it seems reasonable that it is and, indeed, our simulations suggest that, even for $N\ll K$, if a finite $\beta^{\star}$ exists, it is unique.}\footnote{The source code of the implementations in Python is available at https://github.com/angelopiga/info-metric-estimation/.}.

\section{Results}\label{sec.res}

We test our method in a variety of scenarios and compare the results with the main alternative available estimators, the NSB~\cite{nemenman2001,nemenman2004} and the Hausser-Strimmer (HS)~\cite{hausser}. In our experiments, we generate synthetic target distributions and sample multinomial counts $\{n_i\}$ from those distributions. We fix $K = 1000$ and generate samples of increasing size $N = 20, \ldots , 10000$. After calculating $\beta^{\star}$ from~\eqref{eq.result}, we estimate the Shannon entropy $S$ and the Kullback-Leibler divergence $\kl $.
For each case, we repeat this procedure $1000$ times; we always report averages over these repetitions \footnote{Averaging on multiple runs is preferable in order to highlight the scaling behaviors of the estimators while mitigating the effects of outliers (for example, very singular distributions or samples).}. 

As target distributions (see Fig.~\ref{fig.dist} as reference) we consider categorical distributions that are both \textit{typical} in the Dirichlet prior (that is, they are generated by a symmetric Dirichlet prior; we use several values of concentration parameter $\beta = 0.01, 1, 10$) and \textit{atypical} in the Dirichlet prior (that is, they cannot be attributed to or have a negligible probability of being generated from a symmetric Dirichlet prior). Among the latter, we consider: (i) distributions with added structural zeroes (that is, we sample from a symmetric Dirichlet prior with a given $\beta$, but half of the categories are then forced to have zero probability) \footnote{This scenario corresponds to an experiment in which some states are not observable.}; (ii) Zipf's distribution, ubiquitous in nature, in biological as well as social systems~\cite{newman2005power}, characterized by probabilities $\rho_i\propto i^{-a}$, with a exponent $a\ge 1$; (iii) Bimodal distributions, which represent, for example, the degree distributions of core-periphery complex networks~\cite{coreper} \footnote{In Refs.~\cite{nemenman2001,nemenman2004} a rigorous definition of atypicity is provided, related to the shape of the tails of a Zipf's distribution.}.

% --------------------------------------------------------------
\subsection{Shannon entropy}

To estimate the posterior $p(S|\bn)$ of the Shannon entropy we use the exact formulas of its moments (derived in Refs.~\cite{wolpert1,wolpert2} and later refined in Ref.~\cite{archercountable}) with the estimated values of $\beta^{\star}$. The first moment is given by
\begin{equation}\label{eq.wwmean}
\begin{split}
\mathbb{E}[S|\bn,\beta] &=  \int d\brho \; S(\brho|\beta) \; p(\brho |\bn) \\
&= \psi_0(N+K\beta+1) \\
&- \sum_{i=1}^K \frac{n_i+\beta}{N+K\beta}\,\psi_0(n_i+\beta+1)\;.
\end{split}
\end{equation}
%where $\psi_i(x) = (d/dx)^{i+1}\,\log\Gamma(x)$ are the polygamma functions.
In Appendix~\ref{ap:moments} we also show the expression of the standard deviation.

\begin{figure}[t]
    %  \centering
     %\begin{subfigure}[b]{1\columnwidth}
        %  \centering
         \includegraphics[width=\columnwidth]{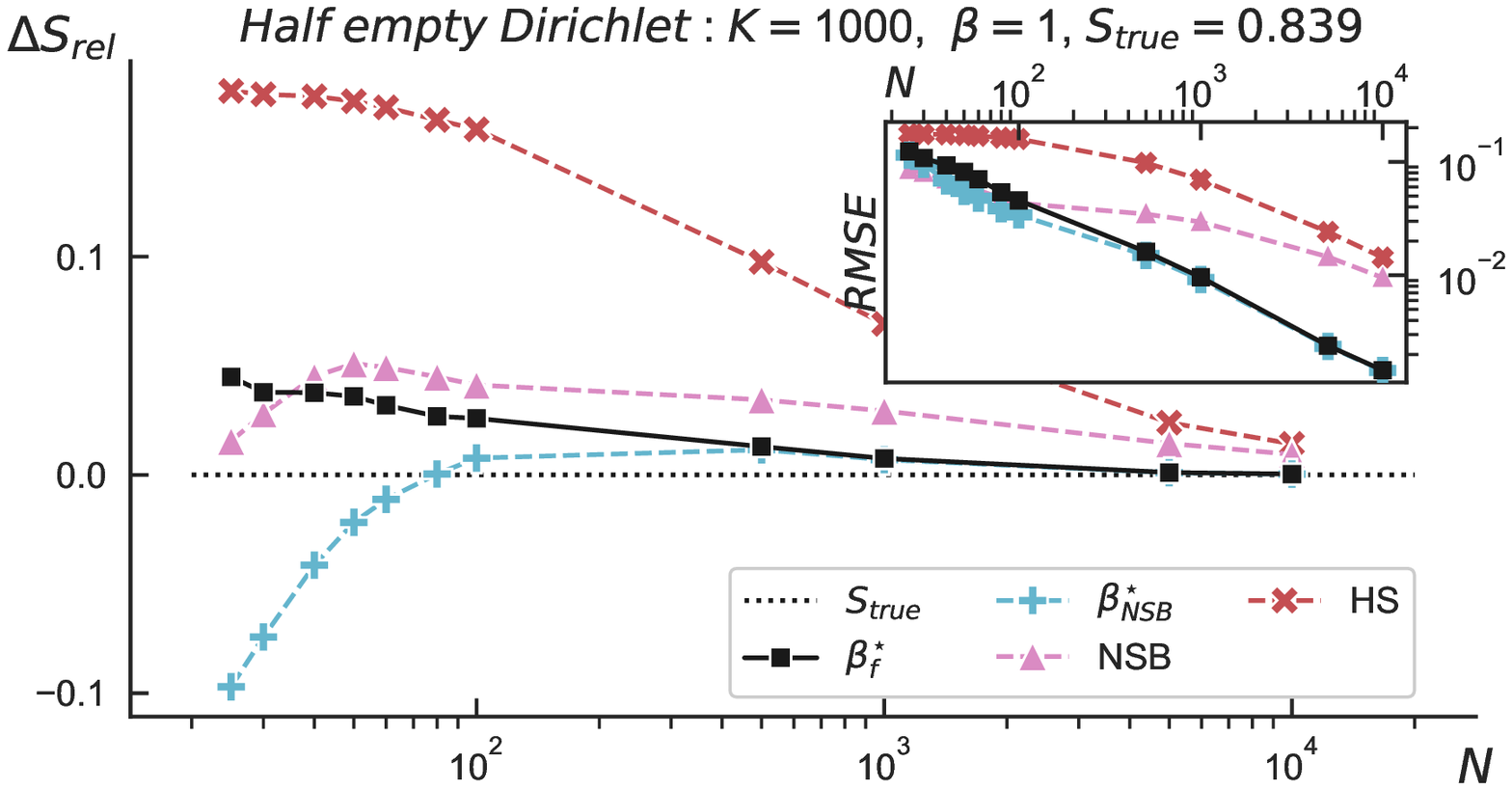}
        %  \caption{$y=x$}
     %\end{subfigure}
     \vfill
     %\begin{subfigure}[b]{1\columnwidth}
        %  \centering
         \includegraphics[width=\columnwidth]{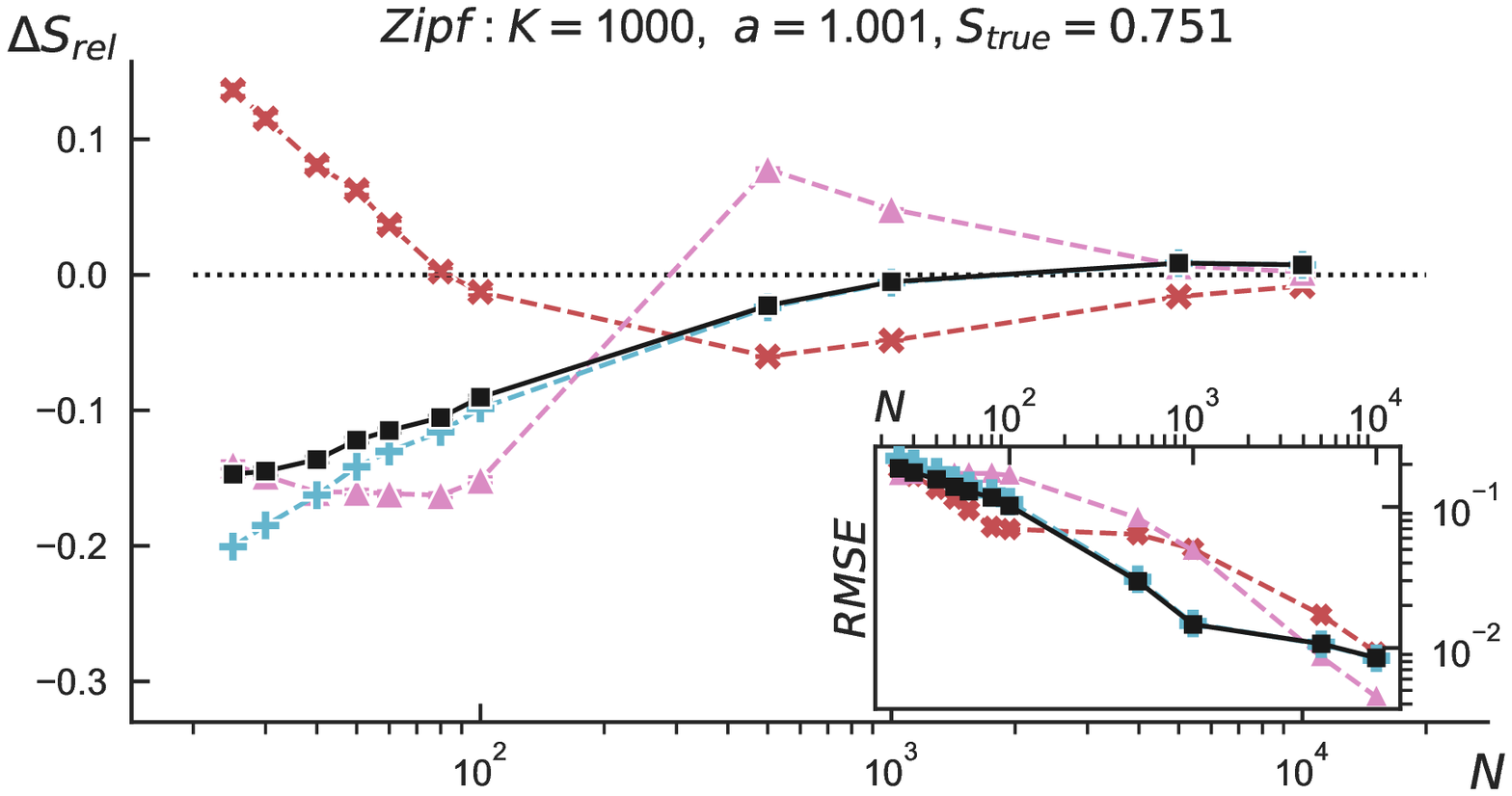}
     %\end{subfigure}
     \vfill
     %\begin{subfigure}[b]{1\columnwidth}
        %  \centering
         \includegraphics[width=\columnwidth]{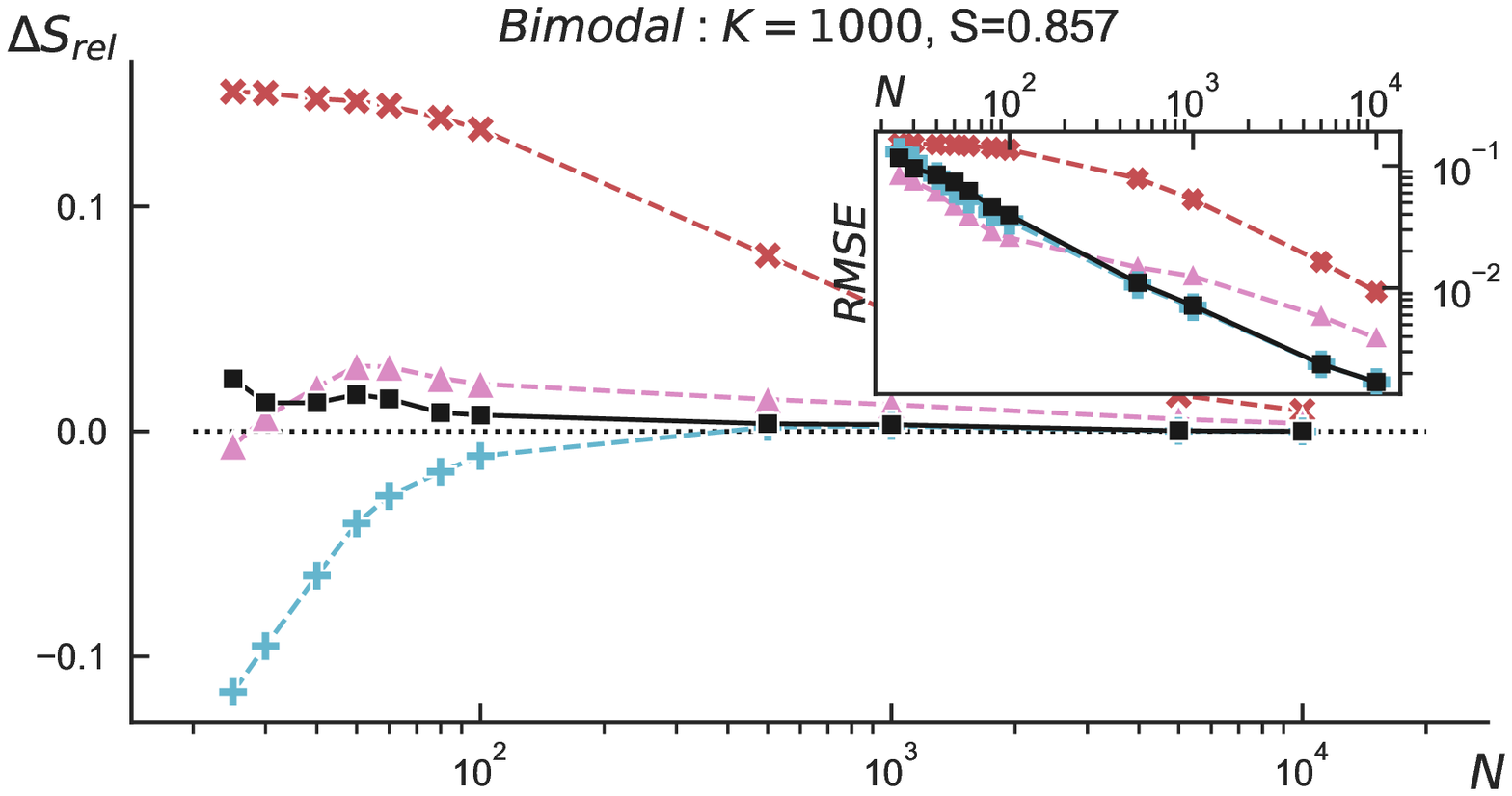}
       %  \caption{}
     %\end{subfigure}
    \caption{Shannon entropy estimation for atypical distributions in a Dirichlet prior: Dirichlet with $\beta=1$ but half bins are set to zeros; Zipf's distribution with exponent $a=1.001$; bimodal distribution. Standard-errors bars of the main plots are smaller then symbols and are not shown.}
    \label{fig:S_atyp}
\end{figure}

In practice, given a dataset $\bn$ we calculate the most probable $\beta^{\star}$ from Eq.~\eqref{eq.result} by assuming either a flat hyperprior, Eq.~\eqref{eq.flat-hyp}, or the NSB hyperprior, Eq.~\eqref{eq.nsb-hyp}. Then, we compute the required moments of the Shannon entropy; we indicate the estimated values of the Shannon entropy as $S(\beta^{\star}_{\rm flat})$ and $S(\beta^{\star}_{\rm NSB})$, respectively. In Figs.~\ref{fig:S_typ} and \ref{fig:S_atyp}, we show that our estimator with a flat hyperprior is the most accurate estimator overall. In particular, $S(\beta^{\star}_{\rm flat})$ is consistently more accurate than the NSB estimator, except in the deep sparse regime $N<30$ of two of the distributions atypical in the Dirichlet prior, where it is comparable but slightly less accurate. The Bayesian estimators also behave better than the HS estimator $S_{\rm HS}$ except for very uniform distributions sampled from the Dirichlet prior with $\beta=10$. Overall, the $S(\beta^{\star}_{\rm flat})$ has little bias often even in the very sparse regime and for distributions atypical in the Dirichlet prior. 
It is also interesting to note that both $S(\beta^{\star}_{\rm flat})$ and $S(\beta^{\star}_{\rm NSB})$ have a more regular scaling behavior, in the convergence toward the true values as $N$ increases, in particular when compared with NSB and HS for Zipf's distribution.

We also analyze the variability of the Shannon entropy estimates, as measured by the root mean squared error (insets in Figs.~\ref{fig:S_typ} and \ref{fig:S_atyp}). This analysis reveals that, besides having less bias, the $S(\beta^{\star}_{\rm flat})$ estimator has a variability that is typically comparable to or smaller than the other estimators. It is also worth noting that, differently from Bayesian estimators, for which all the moments can be estimated also from a single sample, the HS estimator is limited to a point estimate of the mean value of Shannon entropy.

Note that, contrary to what one may expect, $S_{\rm NSB}$ differs from our estimate $S(\beta^{\star}_{\rm NSB})$ in that the latter is always smaller for small samples. This happens because the NSB hyperprior~\eqref{eq.nsb-hyp} is a positive monotonically-decreasing function that assigns higher probabilities to smaller $\beta$'s, while the Shannon entropy of distributions sampled from a symmetric Dirichlet is a monotonically-increasing function of $\beta$. However, it is not the same estimating $\beta^{\star}$ with the NSB hyperprior and then plugging it in~\eqref{eq.wwmean} or directly estimating the Shannon entropy with the NSB prior~\eqref{eq.nsbprior}; the latter in fact provides better results. On the other side, $S(\beta^{\star}_{\rm flat})$ and $S_{\rm NSB}$ should not substantially differ, being based on the same first principles of estimation. 
The differences are attributable to the numerical and computational difficulties in implementing the NSB approach that required both a fine discretization over $\beta$ and solving as many equations~\eqref{eq.wwmean} as $\beta$'s, which have to be finally integrated with weights given by the hyperprior~\eqref{eq.nsb-hyp}, in contrast with our approach, which needs solving just Eq.~\eqref{eq.result} and  Eq.~\eqref{eq.wwmean} once.

% Comparing our approach with the NSB, we note that the amount of numerical calculations is highly reduced

% -----------------------------------------------------------------
\subsection{Kullback-Leibler divergence}

Regarding the Kullback-Leibler divergence $\kl$, there are no exact formulas for the moments of the posterior distribution $p(\kl|\bn)$. Therefore, we have to rely on a point estimate of the mean by first estimating the distributions via Laplace's formula~\eqref{eq.laplace} with the inferred $\beta^{\star}$ and then plugging these values into expression~\eqref{eq.kl}. The flat hyperprior in Eq.~\eqref{eq.flat-hyp} is the only reasonable one to estimate $\beta^{\star}$ in this case, since the NSB prior (Eq.~\eqref{eq.nsb-hyp}) can only be justified for the Shannon entropy.
\begin{figure}[t!]
    %\centering
    %\begin{subfigure}[b]{1\columnwidth}
    %\centering
    \includegraphics[width=\columnwidth]{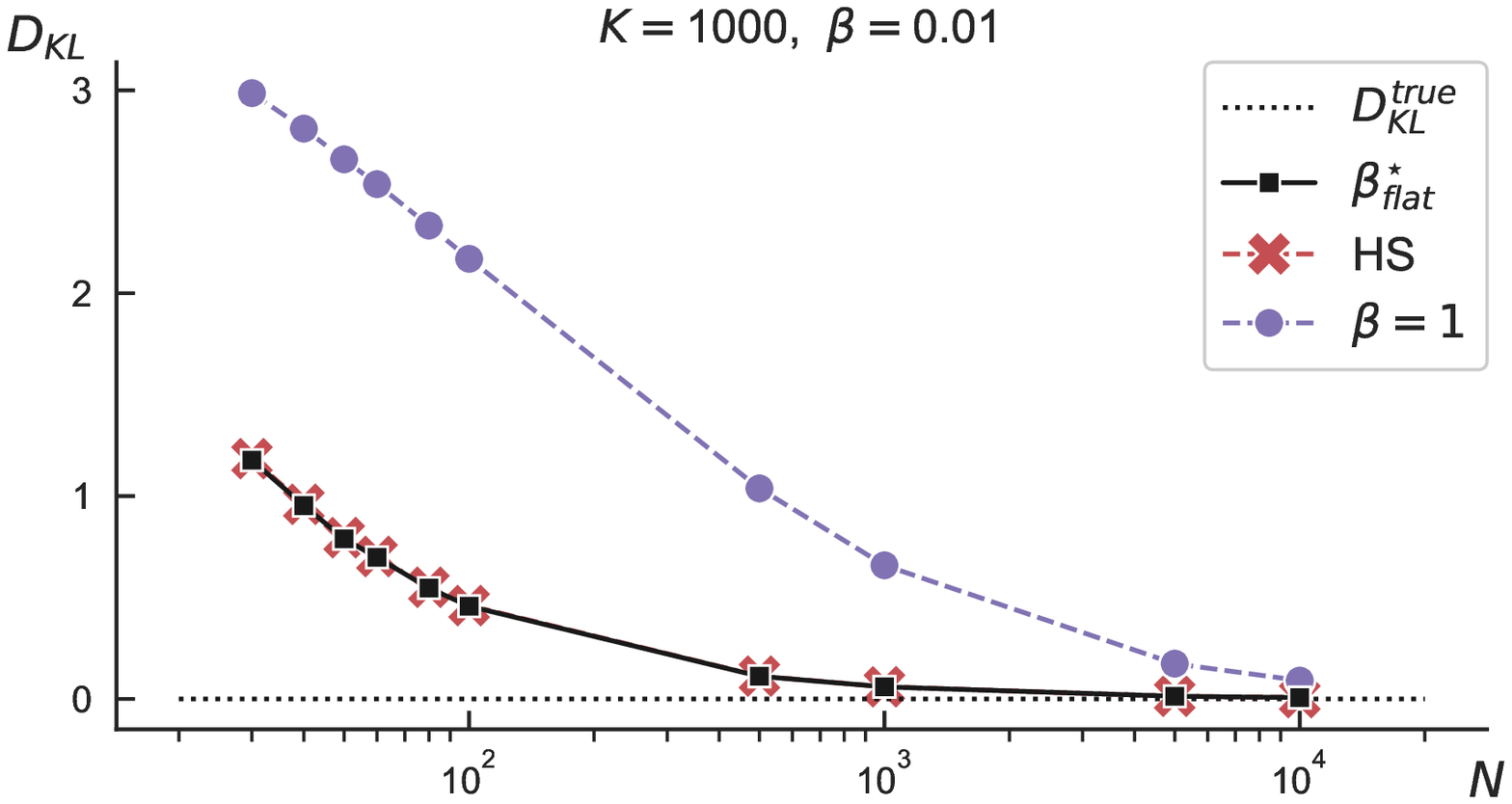}
     %\end{subfigure}
     \vfill
    \includegraphics[width=\columnwidth]{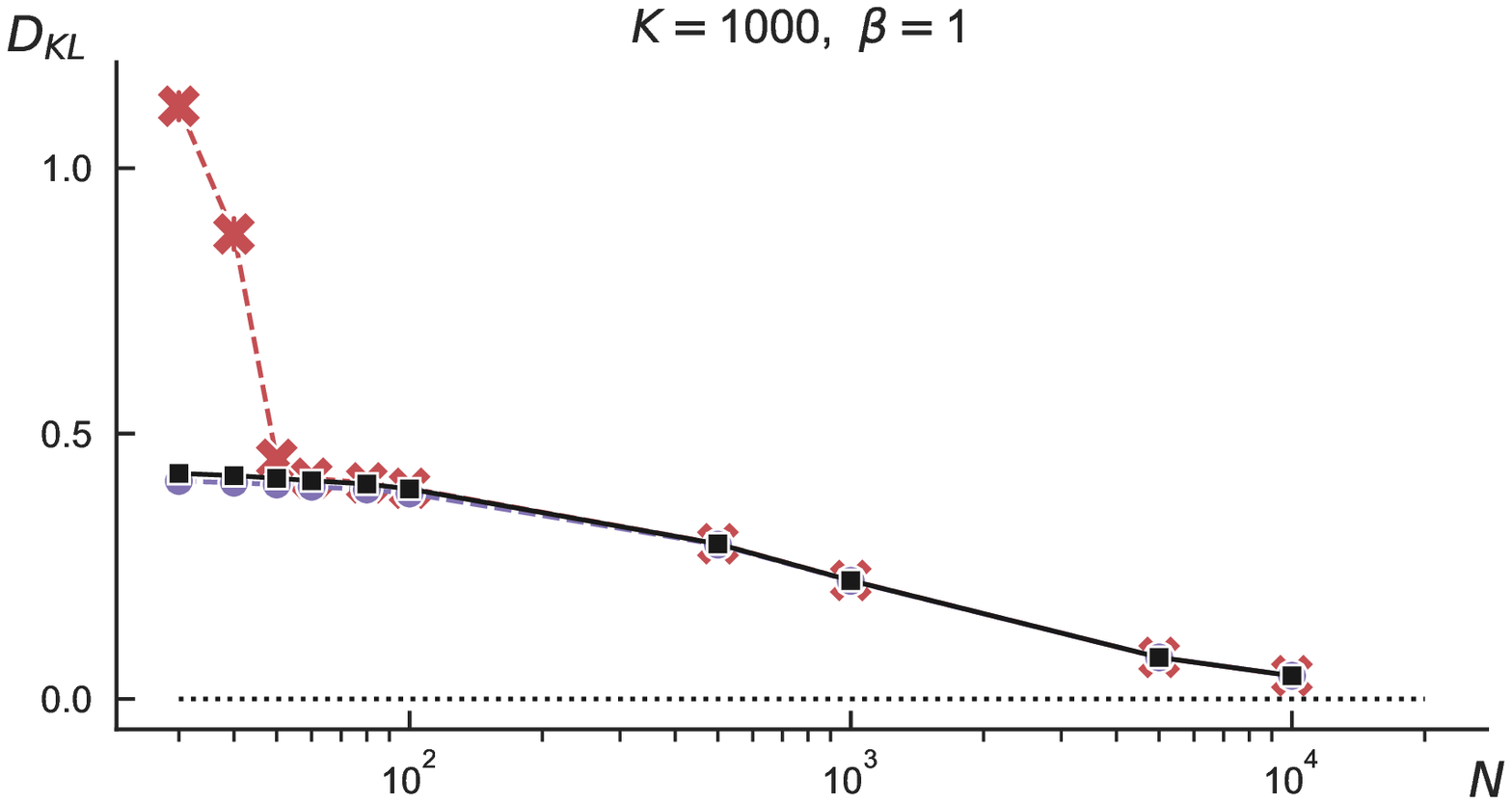}
     \vfill
    \includegraphics[width=\columnwidth]{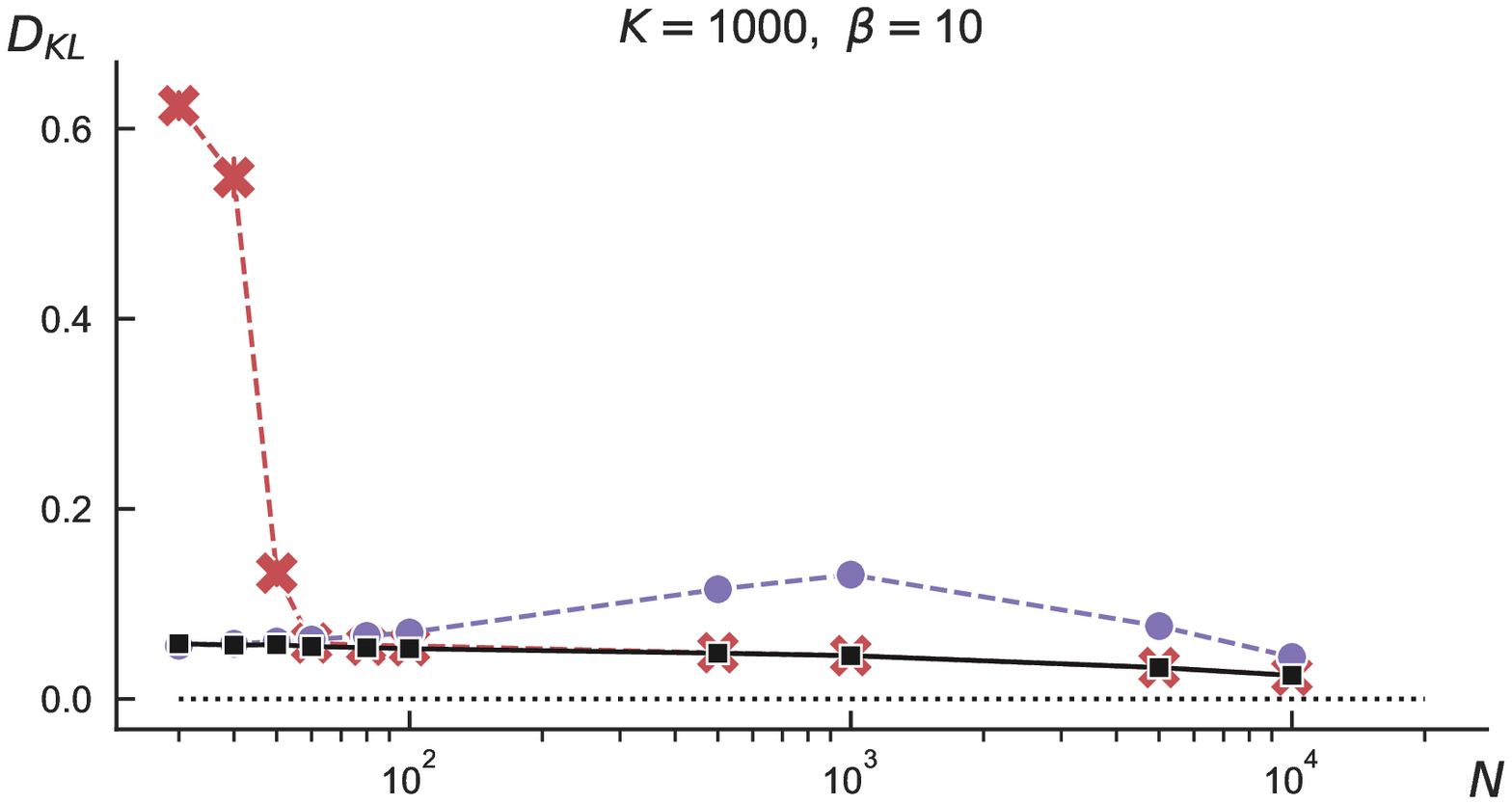} 
    \caption{Kullback-Leibler estimation for distributions typical in a Dirichlet prior, for $\beta=0.01, 1, 101$ and sample size $N = 25\ldots 10000$. Each point corresponds to an average over $1000$ samples. Here, $D_{KL}$ is taken between a given distribution and a second one estimated from a sampling of the former; the target value is therefore $D_{KL}^{true} = 0$.  Black squares: our estimator, that is Laplace's formula with $\beta^{\star}_{\rm flat}$ estimated from a flat hyperprior. Red crosses: Hausser-Strimmer estimator. Purple circles: Laplace's estimator for uniform prior $\beta=1$. Standard-errors bars are often smaller then symbols and are not visible.}
    \label{fig:KL_typ}
\end{figure}

We compare the results with Laplace's estimator~\eqref{eq.laplace} with $\beta = 1$ and with the HS estimator, since both have the same desirable property of assigning non-null probabilities to unobserved states ($n_i=0$) and are suitable estimators for computing $\kl$. Indeed, $\beta = 1$ in Laplace's formula is a common choice and amounts to assigning the same probability to all possible distributions. We test the estimators in a scenario typical in machine learning and variational inference, in which one wants to minimize the $\kl$ between a complex, target distribution and some model approximation. Here, after generating a synthetic discrete distribution $\brho$, we measure the $\kl (\brho ; \hat{\brho})$, where $\hat{\brho}$ is the distribution estimated from counts; hence a good estimator should make $\kl $ as small as possible.  

\begin{figure}[tb]
    %  \centering
     %\begin{subfigure}[b]{1\columnwidth}
        %  \centering
         \includegraphics[width=\columnwidth]{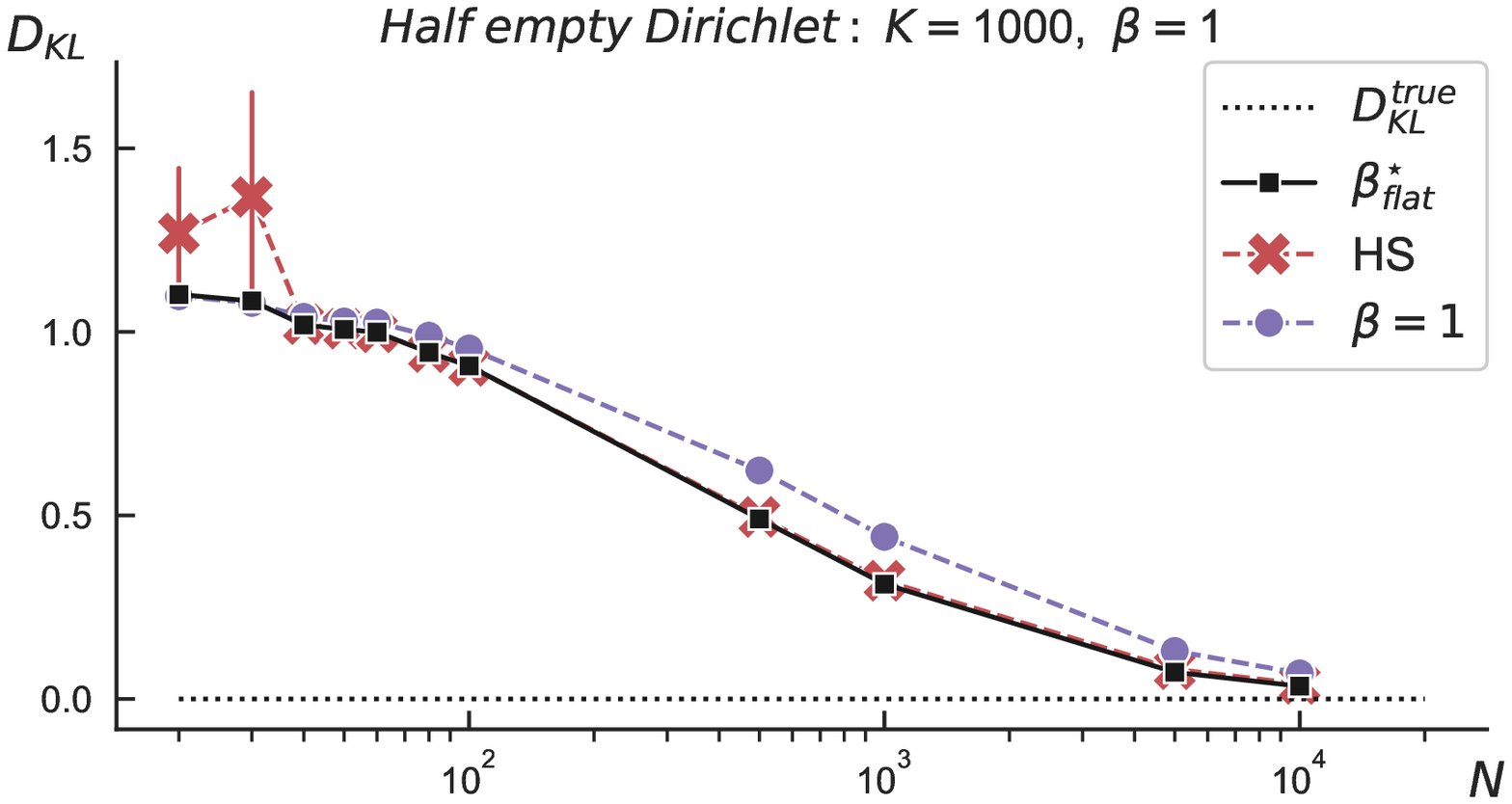}
        %  \caption{$y=x$}
     %\end{subfigure}
     \vfill
     %\begin{subfigure}[b]{1\columnwidth}
        %  \centering
         \includegraphics[width=\columnwidth]{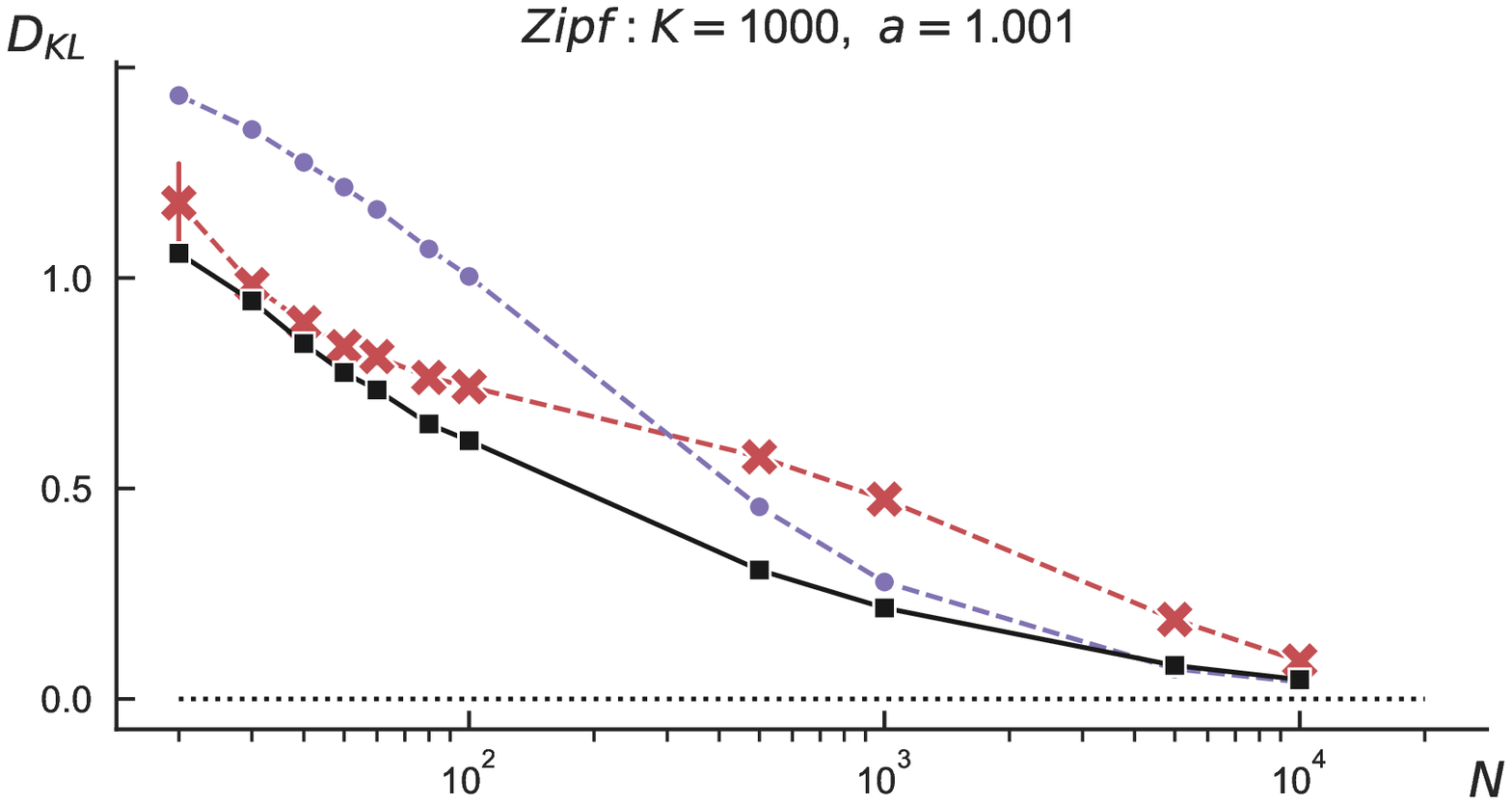}
     %\end{subfigure}
     \vfill
     %\begin{subfigure}[b]{1\columnwidth}
        %  \centering
         \includegraphics[width=\columnwidth]{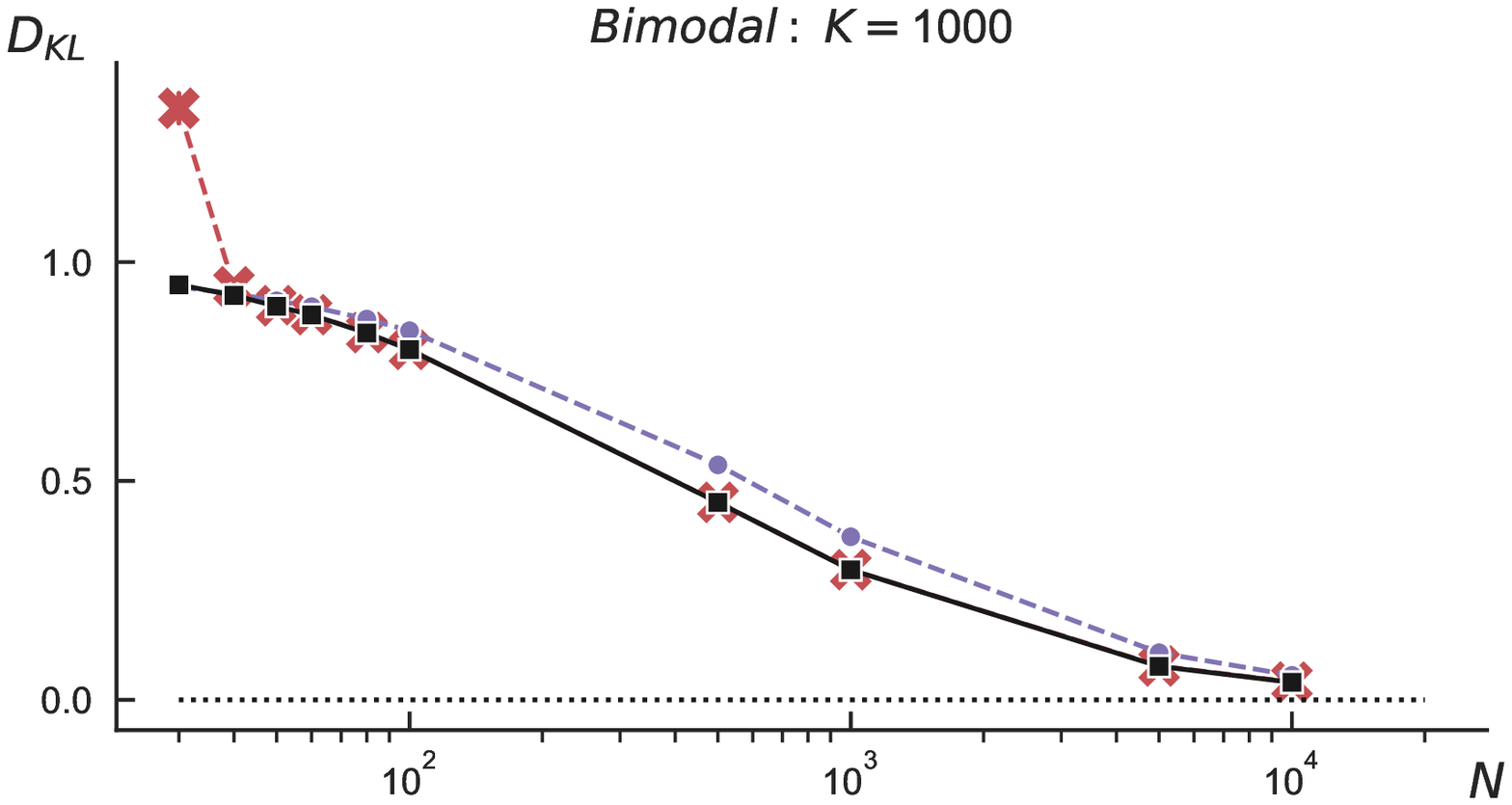}
       %  \caption{}
     %\end{subfigure}
    \caption{Kullback-Leibler divergence estimation for atypical distributions in a Dirichlet prior: Dirichlet with $\beta=1$ but half bins set to zero; Zipf's distribution with exponent $a=1,001$; bimodal distribution. Here, $D_{KL}$ is taken between a given distribution and a second one estimated from a sampling of the former; the target value is therefore $D_{KL}^{true} = 0$. Standard-errors bars are often smaller then symbols and are not visible, except for HS estimator, $N<40$, for half-empty and Zipf's distributions. }
    \label{fig:KL_atyp}
\end{figure}

In Figs.~\ref{fig:KL_typ} and~\ref{fig:KL_atyp}, we show that our estimator and the HS estimator provide similar results, although $\kl (\beta^{\star})$ is more accurate in the very sparse regime $N<50$, and when the target distributions are atypical in the Dirichlet priors, especially in the important case of Zipf's distributions. The estimator based on Laplace's formula wih $\beta=1$ performs generally worse, unless in the trivial case when the target distribution itself was also generated just from a Dirichlet with $\beta=1$. Importantly, in this case in which $\beta=1$ is optimal, our approach provides virtually identical results.

\section{conclusions}

We have addressed the question of how to estimate real distributions and their information-theoretical metrics, such as the Shannon entropy or the Kullback-Leibler divergence, when only a small number of observations are available. The estimation problem in the sparse sampling regime is, theoretically and experimentally, unavoidable in complex systems~\cite{subsampling-rev}. Its rigorous study is then necessary, given the broad use of information-theoretical metrics in physics---especially Shannon entropy and related quantities such as the mutual information---and in data science and machine learning---Kullback-Leibler divergence is at the core of approaches as successful as variational autoencoders or diffusion models, to name just a couple of very prominent examples. 

Very few estimators for Kullback-Leibler divergence have been proposed in the literature; they are more abundant for Shannon entropy. However, in both cases they suffer from limitations. First, many proposed methods fit well for specific contexts but fail in others, because of the implicit presence of {\em ad hoc} assumptions in their derivation. Second, with few exceptions, their implementation requires complex numerical algorithms. Third, they often only provide a point-wise estimation, without any estimation of the error. Avoiding these drawbacks is crucial for developing a research program that is accurate and general, and the laws of conditional probability, in the form of Bayes rules, are what is needed for this purpose~\cite{jaynesbook}. 

In such a framework, the NSB estimator can be still considered the state of the art and has been of inspiration for many subsequent works extending the original method.
Crucial in NSB (and in Bayesian analysis, in general) is the choice of the prior distribution, which explicitly expresses expectations about the generation of the data. The NSB prior is a clever mixture of Dirichlet distributions. In fact, Dirichlet priors were broadly used well before NSB, as they are expressive generative models for discrete distributions; but they gave inconsistent estimations of the Shannon entropy. As pointed out by NSB, the explanation is to be sought in the properties of Dirichlet priors---they are defined by a set of hyperparameters $\beta$ and, as NSB observed, when samples are scarce, the choice of $\beta$ narrowly determines the value of entropy. Since a prior-dependent inference is not useful, and in light of the difficulties in determining the correct {\em a priori} values of $\beta$, NSB circumvented the problem by integrating over all possible values of $\beta$, which ultimately results in the aforementioned mixture of priors. The impressive results of NSB at estimating Shannon entropy come with the shortcomings of requiring a complicated implementation and, above all, renouncing to estimate the probability distribution, which might be necessary for other applications, such as estimating the Kullback-Leibler divergence. 

In this paper, we prove that, whereas mixtures of priors are necessary to accommodate for any possible value of the hyperparameter $\beta$, in practice, considering a single value $\beta^\star$ leads to excellent estimation. As we have shown, this value of the hyperparameter can be found directly, given the few available observations. Far from being a mere technical point, knowing the hyperparameter $\beta^\star$ allows the full specification of the generative model and the estimation of the probability distribution. Importantly, our results still follow from a purely Bayesian framework; more precisely, from a hierarchical probabilistic model, where Bayes' rule is first applied at the higher level for the estimation of the prior parameters. This way, the overall simplicity of the assumptions and transparency of the derivation are preserved. The value of $\beta^\star$ is finally provided by a closed formula (Eq.~\eqref{eq.result}), which is easy to implement and depends only on the vector of observations.
Additionally, as proved by simulations over a broad variety of distributions, our estimators provide results at least as accurate as, and most times more accurate than, state-of-the-art estimators for Shannon entropy and Kullback-Leibler divergence.

Further efforts should be devoted to extend the present approach to  priors other than Dirichlet, for example for data that are likely to follow power law distributions, including Zipf's laws, or when the number of states is unknown, as in Refs.~\cite{chaoshen, archercountable, valiant, wolpertdedeo}. Further efforts will also be necessary to obtain Kullback-Leibler estimators that go beyond the point estimates provided here.

\section{Acknowledgements}
%We acknowledge financial support from Fundaci\'on ``La Caixa'' for the project \textit{Data science of judicial decisions for evidence-based housing policies in Spain} (DaSiHo) (LCF/PR/SR19/52540009) and 
This research was funded by the Social Observatory of the ``la Caixa'' Foundation as part of the project LCF / PR / SR19 / 52540009, by MCIN / AEI / 10.13039 / 501100011033 (Project No. PID2019–106811GB-C31) and by the Government of Catalonia (Project No. 2017SGR-896).

\bibliography{sample}
\bibliographystyle{unsrt}

\appendix
\begin{widetext}
\section{Derivation of results (Eq.~\eqref{eq.result} in main text)}\label{sec.supp}

Let us suppose that we have $K$ different categories (or types of random events) and that we observe $N$ independent random events distributed in the $K$ categories ${\bf n}=\{n_i;\, i=1,\dots,K\}$, with $\sum_i n_i=N$. We also assume that the probabilities of observing counts in each category $\rho_i$ are distributed according to a Dirichlet prior with the same hyper-parameters $\beta$ for all ${\brho=\{\rho_i;\, i=1,\dots,K\}}$, so that
\begin{equation}
    p(\brho|\beta)=\frac{1}{B_K(\beta)} \prod_{i=1}^K \rho_i^{\beta-1}\;, \qquad B_K(\beta)=\frac{\Gamma(\beta)^{K}}{\Gamma(\beta K)}.
    \label{eq.prior_supp}
\end{equation}

Our goal is to compute the most likely value of $\beta$ given the observed counts $\{n_i\}$. To that end, we need to compute the conditional probability $p(\beta|{\bf n})$. We can do this by marginalizing over the possible combinations of $\boldsymbol{\rho}=\{\rho_i\}$ as follows:
\begin{equation}
    p(\beta|\bn) = \frac{p(\beta)}{p(\bn)}p(\bn|\beta)\;, \qquad p(\bn|\beta)=\int d\brho\, p(\bn|\beta,\brho)p(\brho|\beta)~.
    \label{eq.marginal_supp}
\end{equation}
Since the probability of observing an event in category $i$ is $\rho_i$, the probability of observing $n_i$ events of type $i$ is $\rho_i^{n_i}$. Therefore, for the integral in Eq.~\eqref{eq.marginal_supp} we have that
\begin{equation}
    p(\bn|\beta,\brho)=\prod_{i=1}^K\rho_i^{n_i}\;,
\end{equation}
so that
\begin{equation}
p(\bn|\beta)=\frac{1}{B_K(\beta)}\int d\brho\,\prod_{i=1}^K\rho_i^{n_i+\beta-1},
\end{equation}
where we have used Eq.~\eqref{eq.prior_supp} for $p(\brho|\beta)$ and the integral is over the simplex that satisfies the condition $\sum_{i=1} \rho_i=1$.

To perform the integrals above we first evaluate the normalization condition for $\rho_k=1-R(K-1)$ with $R_{K-1}=\sum_{i=1}^{K-1}\rho_i$ so that for $\rho_{k-1}$ we have the following integral:
\begin{equation}
     I_{K-1}= \int_0^{1-R_{K-2}}d\rho_{K-1}\, \rho_{K-1}^{n_{K-1}+\beta-1}\left(1-\rho_{k-1}-R_{K-2}\right)^{n_K+\beta-1}.
    \label{eq.rhok-1}
\end{equation}
To evaluate this integral we use the fact that 
\begin{equation}
    \int_0^{(1-R)}dx\,x^a (1-x-R)^b = \frac{\Gamma(a+1)\Gamma(b+1)}{\Gamma(a+b+2)} (1-R)^{a+b+1}\qquad {\rm if} \quad {\rm Re}(R)<1 \quad {\rm and}\quad {\rm Im}(R)=0
    \label{eq.gamma}
\end{equation}
so that
\begin{equation}
     I_{K-1}=\frac{\Gamma(n_{K-1}+\beta)\Gamma(n_K+\beta)}{\Gamma(n_k+n_{K-1}+2\beta)} (1-R_{K-2})^{n_K + n_{K-1} + 2\beta-1}
    \label{eq.rhok-1s}~~
\end{equation}
Which gives for $\rho_{K-2}$ the following integral:
\begin{eqnarray}
    I_{K-2}&=& \int_0^{1-R_{K-3}}d\rho_{K-2}\;\rho_{K-2}^{n_{K-2}+\beta-1}\left(1-\rho_{K-2}-R_{K-3}\right)^{n_K+N_{K-1}+2\beta-1}\\
    &=&  \frac{\Gamma(n_{K-2}+\beta)\Gamma(n_K+n_{K-1}+2\beta)}{\Gamma(n_k+n_{K-1}+n_{k-2}+3\beta)} (1-R_{K-3})^{n_K + n_{K-1} +n_{K-2}+ 3\beta-1}
    \label{eq.rhok-2}
\end{eqnarray}
which have evaluated using Eq.~\eqref{eq.gamma}.
If we do this for all $\brho$ we end up having
\begin{equation}
    \int d\brho\,\prod_i\rho_i^{n_i+\beta-1}= \prod_{i=1}^{K} I_i=\frac{\prod_{i=1}^K \Gamma(n_i+\beta)}{\Gamma(N+K\beta)}\;.
\label{eq.full}    
\end{equation}
Thus, we obtain the following expression for $p(\bn|\beta)$
\begin{equation}
    p(\bn|\beta)=\frac{1}{B_K(\beta)}\frac{\prod_i \Gamma(n_i+\beta)}{\Gamma(N+K\beta)}=\frac{\Gamma(K\beta)}{\Gamma(\beta )^{K}}\frac{\prod_i \Gamma(n_i+\beta)}{\Gamma(N+K\beta)}
    \label{eq.posterior_supp}
\end{equation}
Our goal is to find $\beta^{\star}$ that maximizes $p(\beta|\bn)= \frac{p(\beta)}{p(\bn)}p(\bn|\beta)$. To that end we take the derivative of $\log p(\beta|\bn)$,
\begin{equation}
    \log p(\beta|\bn)= \log \Gamma(K\beta) - K \log \Gamma(\beta) + \sum_i \log \Gamma(n_i+\beta) - \log \Gamma(N+K\beta)+\log p(\beta)-\log p(\bn)
    \label{eq.logp}
\end{equation}
so that $\beta^{\star}$ is the one that satisfies the condition:
\begin{equation}
    \left.\frac{d \log p(\beta|\bn)}{d\beta}\right|_{\beta=\beta^{\star}} = 0 ~.
    \label{eq.dlogp}
\end{equation}
To evaluate this equation we use the following definitions and properties of the log Gamma function:
\begin{eqnarray}
    {\rm 1.}\quad &  \left(\frac{d}{dx}\right)^{m+1} \log \Gamma(x)=\psi_m(x)\\
    {\rm 2.}\quad & \psi_0(x+n) = \sum_{m=0}^{n-1}\frac{1}{x+m}+\psi(x)~.
\end{eqnarray}
Using the expressions above and the consideration that $p(\beta)={\rm const.}$ we obtain that:
\begin{eqnarray}\label{eq.dlogpfinal}
    \frac{d \log p(\beta|\bn)}{d\beta} & = & K \psi_0(K\beta)-K\psi_0(\beta)+\sum_i\psi_0(n_i+\beta)-K\psi_0(N+K\beta)\\
    &=&\sum_{i=1}^{K}\sum_{m=0}^{n_i-1} \frac{1}{m+\beta} -\sum_{m=0}^{N-1}\frac{K}{m+K\beta}
\end{eqnarray}
Therefore the condition that gives $\beta^{\star}$ is
\begin{equation}
    \sum_{i=1}^{K}\sum_{m=0}^{n_i-1} \frac{1}{m+\beta^{\star}}-\sum_{m=0}^{N-1}\frac{K}{m+K\beta^{\star}}=0 \;,
\end{equation}
that is, the Eq.~\eqref{eq.result} in main text for uniform hyperprior~\eqref{eq.flat-hyp}.
If instead we consider a prior for beta that results in a close-to-uniform distribution of Shannon entropy such as in Nemenman et al.~\cite{nemenman2001,nemenman2004} then
\begin{equation}
 p_{NSB}(\beta)=\frac{d \overline{S}}{d\beta} ~,   
\end{equation}
 with $\overline{S} = \mathbb{E}[S|n_i=0,\beta] = \psi_0(K\beta+1)-\psi_0(\beta+1)$, the average entropy of the distributions generated from a Dirichlet prior $p(\brho|\beta)$. Note that this prior is already normalized since $\int_0^{\infty}d\overline{S}/d\beta d\beta=\overline{S(\infty;K)}-\overline{S(0;K)}=1$. The derivative of the logarithm of this prior with respect to $\beta$ is then
\begin{equation*}
\frac{d \log p_{NSB}(\beta)}{d\beta}= \frac{1}{p_{NSB}(\beta)}\frac{d p_{NSB}(\beta)}{d\beta}=\frac{1}{\frac{d\overline{S}}{d\beta}} \frac{d^2 \overline{S}}{d \beta^2}=\frac{K^2\psi_2(k\beta+1)-\psi_2(\beta+1)}{K\psi_1(k\beta+1)-\psi_1(\beta+1)} ~,   
\end{equation*}
which is the formula~\eqref{eq.result} in main text. The condition of the $\beta^{\star}$ that maximizes $p(\beta|n)$ is in this case:
\begin{eqnarray*}
    \frac{d \log p(\beta|\bn)}{d\beta}& = &K\psi_0(K\beta)-K\psi_0(\beta)+\sum_i\psi_0(n_i+\beta)-K\psi_0(N+K\beta)\nonumber+ \frac{1}{\frac{d\overline{S}}{d\beta}} \frac{d^2 \overline{S}}{d \beta^2}=\\
   &=&\sum_{i=1}^{K}\sum_{m=0}^{n_i-1} \frac{1}{m+\beta^{\star}}-\sum_{m=0}^{N-1}\frac{K}{m+K\beta^{\star}}\nonumber + \frac{K^2\psi_2(k\beta^{\star}+1)-\psi_2(\beta^{\star}+1)}{K\psi_1(k\beta^{\star}+1)-\psi_1(\beta^{\star}+1)}=0~.
    \label{eq.dlogpNMB}
\end{eqnarray*}

\section{Analytical moments of the Shannon entropy posterior} \label{ap:moments}

In the specific case of $S(\brho)$, instead of solving $p\left(\mathcal{F}|\bn \right) = \int d\brho \; \delta \left( \mathcal{F} - \mathcal{F}(\brho)\right) p(\brho |\bn)$ (Eq.~\eqref{eq.function} in main text) directly, it is possible to obtain closed form expression for all the moments of the posterior \cite{wolpert1,wolpert2,archercountable}. Here we report the first two, the mean
\begin{equation}
\begin{split}
\mathbb{E}[S|\bn,\beta] =  \int d\brho \; S(\brho|\beta) \; p(\brho |\bn)\; =\; \psi_0(N+K\beta+1) - \sum_{i=1}^K \frac{n_i+\beta}{N+K\beta}\,\psi_0(n_i+\beta+1)\;,
\end{split}
\end{equation}
and the second moment
\begin{equation}\label{eq.wwstd}
\begin{split}
\mathbb{E}[S^2|\bn,\beta] &= \int d\brho \; S(\brho|\beta)^2 \; p(\brho |\bn) 
= \sum_{i\neq j}^K \frac{(n_i+\beta)\,(n_j+\beta)}{(N+K\beta+1)\,(N+K\beta)}\,I_{i,j} + \sum_{i=1}^K \frac{(n_i+\beta+1)\,(n_i+\beta)}{(N+K\beta+1)\,(N+K\beta)}\,J_i\;,
\end{split}
\end{equation}
with
\begin{equation*}
\begin{split}
    I_{i,j} &= \Big(\psi_0(n_i+\beta+1) - \psi_0(N+K\beta+2) \Big)\,  \cdot \Big(\psi_0(n_j+\beta+1) - \psi_0(N+K\beta+2) \Big)
     - \psi_1(N+K\beta+2) \;; \\
\end{split}
\end{equation*}
\begin{equation}
\begin{split}
    J_i &= \Big(\psi_0(n_i+\beta+2) - \psi_0(N+K\beta+2) \Big)^2\, + \psi_1(n_i+\beta+2) - \psi_1(N+K\beta+2)\;;
\end{split}
\end{equation}
from which the standard deviation is in turn calculated as the square root of the variance  $\text{Var}(S|\bn,\beta) = \mathbb{E}[S^2|\bn,\beta]-\mathbb{E}[S|\bn,\beta]^2$.

\end{widetext}

\end{document}